\documentclass[a4paper,11pt]{article}
\pdfoutput=1 
\usepackage{jcappub} 
\usepackage{aas_macros}
\usepackage[T1]{fontenc} 
\usepackage[utf8]{inputenc}
\usepackage[dvipsnames,table]{xcolor}

\usepackage{mwe,bm,verbatim,cprotect,xcolor,xspace}
\usepackage{orcidlink}
\usepackage{subfig,csquotes,float, ulem}
\usepackage{bbold}
\usepackage[export]{adjustbox}
\usepackage{multirow}

\newcommand{\code}[1]{{\texttt{#1}}}
\renewcommand{\emph}[1]{\textit{#1}}

\def\hinvmpc{h^{-1}\,{\rm Mpc}}
\def\hmpcinv{h\,{\rm Mpc}^{-1}}
\def\Omg{\Omega^{\rm MG}}
\def\OmgZero{\Omega^{\rm MG,0}}

\def\chifit{\chi^2_{\rm fit}}
\def\chicurrent{\chi^2_{\rm current\, data}}
\def\gone{\gamma^{\mathrm{MG},0}_1}
\def\gtwo{\gamma^{\mathrm{MG},0}_2}
\def\gthree{\gamma^{\mathrm{MG},0}_3}

\newcommand{\LCDM}{$\Lambda$CDM}

\def\fs{f\sigma_8}

\newcommand{\refeq}[1]{Eq.~(\ref{eq:#1})}     
\newcommand{\refeqs}[2]{Eqs.~(\ref{eq:#1})--(\ref{eq:#2})}     
     
\newcommand{\reffig}[1]{Fig.~\ref{fig:#1}}

\newcommand{\reftab}[1]{Tab.~\ref{table:#1}}     
\newcommand{\refsec}[1]{Sec.~\ref{sec:#1}}
\newcommand{\refapp}[1]{App.~\ref{app:#1}}

\definecolor{WildStrawberry}{HTML}{EE2967}

\definecolor{RoyalBlue}{rgb}{0.25,.41,.88}

\title{Sweeping Horndeski Canvas: New Growth-Rate Parameterization for Modified-Gravity Theories}
\author[a,b,1]{Yuewei Wen\orcidlink{0000-0003-2579-7039}\note{Co-leads.}}
\author[a,b,1]{Nhat-Minh Nguyen\orcidlink{0000-0002-2542-7233}}
\author[a,b]{Dragan Huterer\orcidlink{0000-0001-6558-0112}}
\affiliation[a]{Leinweber Center for Theoretical Physics, University of Michigan, 
450 Church St, Ann Arbor, MI 48109-1040}
\affiliation[b]{Department of Physics, University of Michigan, 450 Church St, Ann Arbor, MI 48109-1040}

\emailAdd{ywwen@umich.edu}
\emailAdd{nguyenmn@umich.edu}
\emailAdd{huterer@umich.edu}

\abstract{We propose and numerically validate a new fitting formula that is sufficiently accurate to model the growth of structure in Horndeski theories of modified gravity for upcoming Stage IV and V large-scale structure surveys. Based on an analysis of more than 18,000 Horndeski models and adopting the popular parameterization of the growth rate $f(z) = \Omega_{M}(z)^{\gamma}$, we generalize the constant growth index $\gamma$ to a two-parameter redshift-dependent quantity, $\gamma(z)$, that more accurately fits these models.  We demonstrate that the functional form $\gamma(z)=\gamma_0+\gamma_1z^2 / (1+z)$  improves the median $\chi^2$ of the fit to viable Horndeski models by a factor of $\sim40$ relative to that of a constant $\gamma$, and is sufficient to obtain unbiased results even for precise measurements expected in Stage IV and V surveys. Finally, we constrain the parameters of the new fitting formula using current cosmological data.}

\keywords{growth of structure, growth rate, growth index, modified gravity, Horndeski theory, large-scale structure}

\begin{document}

\date{\today}

\maketitle
\flushbottom
\clearpage

\section{Introduction}

Over $\sim$13.7 billions years of cosmic evolution, the tiny primordial fluctuations seeded during inflation --- under gravitational interaction --- evolve into the large-scale structure observed and measured by galaxy surveys today. The temporal growth of cosmic structure has a rich and well-understood behavior in different epochs: it is robust in the matter-dominated era, but suppressed at late times, especially following the onset of dark energy. The clustering of galaxies, the weak gravitational lensing of distant background galaxies, and arguably the abundance of galaxy clusters as a function of their redshifts and mass proxies have all established themselves as powerful probes of structure growth. These measurements then translate into constraints on models of dark matter and dark energy or modified gravity (see, e.g. \cite{Hou:2023review} for a recent, general review). 

In the linear regime (corresponding to scales $k\lesssim 0.1\hmpcinv$ today), the growth of density fluctuations is described by the linear growth function $D(a)\equiv\delta(a)/\delta(a=1)$. From it, we can define the \emph{growth rate} as
\begin{equation}
    f(a)\equiv \frac{d\ln D}{d\ln a},
    \label{eq:growth_rate}
\end{equation}
where $a$ is the cosmic scale factor. For the standard, smooth dark-energy with an equation of state $w$, on sub-horizon-scales and in the absence of massive neutrinos, $D$ and $f$ are scale-independent\footnote{See the discussion in the \refapp{scaledep} for more details on the scale-independence in the context of Horndeski models considered in this work.}.  The growth rate can be formally obtained by solving the second-order differential equation which, in standard gravity and on sub-horizon scales, reads $\ddot{D} + 2 H \dot{D} = 4\pi G \rho_M D$. Here $\rho_M$ is the background matter density, $G$ is the Newton constant, and dots are derivatives with respect to time.

In a wide class of cosmological models, the growth rate is well-approximated by a fitting function\footnote{Throughout, we characterize the time evolution interchangeably by the scale factor $a$ or the redshift $z=1/a-1$.}
\begin{equation}
    f(z) = \Omega_M(z)^{\gamma(z)}.
    \label{eq:gamma_parameterization}
\end{equation}
Here, $\Omega_M(z)$ is the time-dependent matter density relative to critical, and the free function $\gamma(z)$ is the so-called \textit{growth index}. The latter has a long history in cosmology, dating back to \cite{Peebles:1980lss,Fry:1985,Lightman_Schechter:1990,Wang:1998gt}, as it describes the growth rate in standard matter-dominated cosmologies.
Ref.~\cite{Linder:2005in} proposed and verified that the growth-index parameterization $\gamma = 0.55+0.02(1+w(z=1))$ fits the true growth for all $w$CDM models to better than 0.2\%, all the way from the matter-dominated era to the present time, for a wide range of $\Omega_M$ values. Moreover, this parameterization also provides a good fit to the growth in some modified-gravity models with, for example, $\gamma\simeq 0.67$ for the DGP models \cite{Linder:2007hg,Gong:2008}.

A fitting formula for the growth rate such as \refeq{gamma_parameterization} is useful for at least two reasons. First, it is easy to implement in cosmological analyses. Second, it is straightforward to test whether the growth agrees with the prediction of a cosmological model (say $\gamma\simeq 0.55$ for \LCDM\ \cite{Nguyen:2023} (see also, e.g. \cite{Alam:2016,Johnson:2016}). Therefore, there has been considerable interest in developing phenomenological formulae for the growth rate and, in particular, investigating their robustness with respect to the choice of the cosmological model.
Analytic works, e.g. \cite{Linder:2007hg,Lee-Ng:2010,Linder:2019,Calderon:2019vog}, have examined the best-fit $\gamma(z)=\gamma_0$ --- the redshift-independent values of $\gamma$ that minimized deviations of \refeq{gamma_parameterization} from the exact growths defined in \refeq{growth_rate} --- in various modified gravity and dark energy models. Further, \cite{Polarski:2008,Gannouji:2008wt,Wu:2009zy,Belloso_2011,Basilakos:2016xob,Basilakos:2019hlb,Khyllep:2019odd,Mirzatuny:2019,Khyllep:2021pcu,Sharma:2021ivo} focused on the redshift evolution of the growth index in $f(R)$ gravity, $f(Q)$ gravity, and interacting dark energy models. The most common redshift-dependent growth-index descriptions are the linear parameterization $\gamma(z) = \gamma_0 + \gamma_1 z$ and the Taylor expansion $\gamma(z) = \gamma_0 + \gamma_1\,z/(1+z)$, the latter of which is motivated by the parameterization $w(a)=w_0+w_a(1-a)$ for the time-dependent equation of state of dark energy $w(a)$ \cite{Cooray:1999da,Linder:2002et}.

Our principal goal in this paper is to explore the accuracy of different $\gamma(z)$ parameterizations within the viable space of Horndeski theory, in the context of future constraints on $f(z)\sigma_8(z)\equiv\fs(z)$ by Stage-IV and Stage-V large-scale structure surveys. These upcoming surveys will yield $\fs(z)$ measurements with small error bars and extend to high redshifts. To that effect, we adopt exact numerical calculations of \refeq{growth_rate} as the baseline, and then fit the growth rate $\fs(z)$ in $\sim$18,000 Horndeski models using a broad set of functional forms. We then compare their goodness of fit to the ground truth in Horndeski models in the context of errors predicted for future surveys, and propose the best new parameterization of $\gamma(z)$. We further demonstrate the utility of the proposed parameterization  by constraining its parameters using current observational data of type Ia supernovae, large-scale structure, and the cosmic microwave background (CMB), and briefly comment on the implications for stress-testing the standard cosmological model.

The rest of this paper is structured as follows. \refsec{theory} reviews the effective field theory framework we exploit to evaluate the growth rate $\fs(z)$ in a given Horndeski model. \refsec{fitting_form} details our procedure to sample Horndeski models and to obtain each model's theoretical prediction on $\fs(z)$.  \refsec{results} discusses various parameterizations of $\gamma(z)$ focusing on their performance in fitting the theory models, and presents current constraints on parameters of the best fitting formula. Finally, \refsec{conclusion} summarizes our analysis.

\section{Horndeski models: theory background  and growth of structure}\label{sec:theory}

We wish to consider the most general class of \LCDM\ extensions for the accelerating universe that is \emph{not} strongly disfavored by current data. We therefore must find an effective way to sample the model space. To this end, we adopt the Effective Field Theory (EFT) formalism for dark energy and modified gravity (henceforth EFTDE) \cite{Gubitosi:2012hu,Bloomfield:2012ff}. Within EFTDE, models with similar properties are established through a grouping of terms in the fundamental Lagrangian such that one can consider a class of models together (for more details, see \cite{Frusciante:2019xia} and references therein).

\subsection{Effective Field Theory approach to Dark Energy}\label{sec:EFTCAMB}

In general, the EFTDE action in unitary gauge can be written as, e.g. \cite{Hu:2013twa,Raveri:2014cka}
\begin{align}
    \label{eq:EFTDE_action}
    S & = \int d^4x\sqrt{-g}\bigg[\frac{1}{2}m_0^2[1+\Omega(t)]R -\Lambda(t)-c(t)g^{00} + \frac{M_2^4(t)}{2}(\delta g^{00})^2 \nonumber \\
    & - \frac{\bar{M}_1^3(t)}{2}\delta K\delta g^{00}
    -\frac{\bar{M}_2^2(t)}{2}\delta K^2  -\frac{\bar{M}_3^2(t)}{2}\delta K_{\nu}^{\,\,\,\mu}\delta K_{\mu}^{\,\,\,\nu} \nonumber \\
    & +\frac{\hat{M}^2(t)}{2}\delta R^{(3)}\delta g^{00}+m_2(t)\partial_ig^{00}\partial^ig^{00}
    +\mathcal{L}_m\bigg],
\end{align}
where $\delta g^{00} = g^{00} +1$ is the perturbation to the time component of the metric, $R^{(3)}$ is the perturbation to the spatial component, and $\delta K_{\mu\nu}$ is the perturbation of the extrinsic curvature. The background evolution depends on three EFTDE functions, $c(t)$, $\Lambda(t)$, and $\Omega(t)$. For any given expansion history, the first two functions, $c(t)$ and $\Lambda(t)$ can be constrained by the Friedmann equations and correspond to energy density and pressure. The effect of modified gravity is parameterized by the third function $\Omega(t)$. The other EFTDE functions in \refeq{EFTDE_action} represent perturbations around the background and correspond to observables that can be compared with observations. Tab. 1 in \cite{Linder:2015rcz} gives a summary of all models that can be represented by the EFT formalism. 

Within EFTDE, we focus on the Horndeski class of models (see, e.g. \cite{Kobayashi:2019hrl} for an in-depth review).
The class of Horndeski theories is the most general scalar-tensor extension of general relativity, including but not limited to quintessence (see \cite{Tsujikawa:2013} for a review) and generalized Brans-Dicke (Jordan Brans-Dicke \cite{Brans_Dicke:1961}, f(R) \cite{Hu_Sawicki:2007}, chameleons \cite{Khoury_Weltman:2004}) models. Moreover, within these scalar-tensor theories, a coupling between the derivative of a scalar field and the Einstein tensor (or the Ricci tensor alone) leads to an accelerated expansion of the cosmic background without demanding a scalar potential \cite{Amendola:1993,Deffayet:2010}.

The Horndeski class is specified by imposing additional constraints on the EFTDE functions that describe perturbations around the background, as follows: 
\begin{align}
     2\hat{M}^2 = \bar{M}_2^2 = -\bar{M}_3^2, \; m_2 = 0.
    \label{eq:Horndeski_constraint}
\end{align}

To evaluate the growth rate in a given Horndeski model and cosmology, we employ the \code{EFTCAMB} framework\footnote{\href{https://github.com/EFTCAMB/EFTCAMB}{github.com/EFTCAMB/EFTCAMB}} \cite{Hu:2013twa,Hu:2014oga,Raveri:2014cka}. \code{EFTCAMB} characterizes a given Horndeski model by seven functions which we parameterize as follows. One aforementioned function, $\Omega(t)$, controls the background evolution.
We henceforth relabel it $\Omg$, for it not to be confused with an energy density parameter.

Inspired by $f(R)$ gravity and (again) following the convention in \cite{Hu:2013twa,Hu:2014oga,Raveri:2014cka}, we further assume $\Omg(t)$ evolves in time as
\begin{equation}
\Omg(a) = \OmgZero a^{s_0}.
\label{eq:Omega_time_evolution}
\end{equation}
For \LCDM, $\Omg(a)=0$. Further, there are six dimensionless, second-order EFTDE functions $\{\gamma^{\mathrm{MG}}_1,\ldots ,\gamma^{\mathrm{MG}}_6\}$ that jointly define the perturbative properties of the model. These functions are related to the perturbation functions in the EFTDE action in \refeq{EFTDE_action} through
\begin{equation}
\begin{aligned}
\gamma^{\mathrm{MG}}_1(t)&=\frac{M_2^4(t)}{m_0^2H_0^2},\,\,\,\,\,\,&&\gamma^{\mathrm{MG}}_2(t)=\frac{\bar{M}_1^3(t)}{m_0^2H_0},\,\,\,\,\,\,\,\,&&&\gamma^{\mathrm{MG}}_3(t)=\frac{\bar{M}_2^2(t)}{m_0^2},\\
\gamma^{\mathrm{MG}}_4(t)&=\frac{\bar{M}_3^2(t)}{m_0^2},\,\,\,\,\,\,\,\,\,\,\,\,\,&&\gamma^{\mathrm{MG}}_5(t)=\frac{\hat{M}^2(t)}{m_0^2},\,\,\,\,\,\,\,\,\,\,\,\,\,\,&&&\gamma^{\mathrm{MG}}_6(t)=\frac{m_2^2(t)}{m_0^2}.
\end{aligned}
\label{eq:gammas_and_Ms}
\end{equation}
We assume that the time evolution of these quantities follows a similar functional form to that of $\Omg(t)$ in \refeq{Omega_time_evolution} above, that is
\begin{equation}
    \gamma^{\mathrm{MG}}_i(a) = \gamma^{\mathrm{MG},0}_i a^{s_i}.
    \label{eq:gamma_time_evolution}
\end{equation}
Note that \refeq{Omega_time_evolution} and \refeq{gamma_time_evolution} implicitly limit the Horndeski theory space accessible in our analysis.

The constraint in \refeq{Horndeski_constraint} corresponds to $2\gamma^{\mathrm{MG}}_5 = \gamma^{\mathrm{MG}}_3 = -\gamma^{\mathrm{MG}}_4$ and $\gamma^{\mathrm{MG}}_6 = 0$.
Therefore, the Horndeski models are fully specified with six EFTDE parameters that control perturbations, $\gamma^{\mathrm{MG},0}_{1,2,3}$ and $s_{1,2,3}$, plus two EFTDE parameters that control the background, $\OmgZero$ and $s_0$.

In this work, we follow the "designer approach" (see e.g.\ \cite{Raveri:2019mxg}) to construct a Horndeski model, where we specify background cosmological parameters and construct the full theoretical model by specifying these standard cosmological parameters that control the expansion rate $H(z)$ and the density perturbations. For the background expansion, we consider a flat \LCDM\ cosmological model, specified by the physical baryon and cold-dark-matter densities ($\Omega_b h^2$ and $\Omega_c h^2$ respectively), and the constant dark-energy equation of state $w=-1$. Our choice of flat geometry implicitly implies that the dark energy density is given as $\Omega_{\rm DE} = 1 - \Omega_M$. Our full model parameter space is therefore
\begin{equation}
    p_i\in \{\Omega_b h^2, \Omega_c h^2, H_0, \Omega^{\rm MG, 0}, \gone, \gtwo, \gthree, s^0, s^1, s^2, s^3\}.
    \label{eq:pars}
\end{equation}

Certain analyses, e.g.\ \cite{Kreisch:2017uet,Noller:2018wyv,Frusciante:2018jzw}, fix $\gamma^{\mathrm{MG}}_3(a)=0$ to enforce that the speed of the propagation of gravitational waves (GW)  be equal to the speed of light. This requirement is motivated by the constraints derived from the binary neutron-star merger events ``observed'' by both GW and optical instruments (see \cite{Ezquiaga_Zumalacarregui:2018} for a review), e.g.\ GW170817 and GRB170817A \cite{Abbott:2017}. To better understand this, consider a simple model-independent parameterization of the speed of propagation of GW \cite{Hou:2023review}
\begin{equation}
c^2_T/c^2=1+\alpha_T(a),
\label{eq:GW_speed}
\end{equation}
where $c^2_T$ and $c^2$ are the squared speeds of GW and of light, respectively. Here $\alpha_T(a)$ quantifies the GW speed's deviation from the speed of light, and can be mapped into the $\gamma_3^{\rm MG}$ function as \cite{Kreisch:2017uet}
\begin{equation}
\alpha_T(a)=-\frac{\gamma^{\mathrm{MG}}_3(a)}{1+\Omg(a)+\gamma^{\mathrm{MG}}_3(a)}.
\label{eq:alpha_T}
\end{equation}
From \refeq{alpha_T}, it is clear that the GW constraint of $-6\times10^{-15}<\alpha_{T,0}<1.4\times10^{-15}$ \cite{Ezquiaga_Zumalacarregui:2018,Abbott:2017} translates into
\begin{equation}
-1.4\times10^{-15}[1+\Omg(a)]<\gamma^{\mathrm{MG}}_3(a)<6\times10^{-15}[1+\Omg(a)],
\label{eq:GW_derived_constraint}
\end{equation}
which is often simply taken to be $\gamma^{\mathrm{MG}}_3(a)=0$ within EFTDE (or $\alpha_T=0$ in general).

In this paper, we do \emph{not} follow the above approach but rather, for full generality, allow for $\gthree\neq0$. This choice certainly merits a justification: \cite{deRham:2018red} pointed out that current LIGO multi-messenger GW events have only been detected at the energy scale close to either the strong coupling scale or the EFT cut-off. They further explicitly showed that, within the EFTDE approach to Horndeski theories, the GW speed is generally a function of energy scale $c_T(k)$ (see their Eq.~(13)), and therefore can still potentially deviate from the speed of light when measured at lower frequencies (see their Fig.~1). Those Horndeski models hence do not necessarily obey the derived constraint in \refeq{GW_derived_constraint}. Future observations of either GW events at a lower frequency, e.g.\ with LISA \cite{Baker:2019nct}, or CMB B-mode polarization \cite{LiteBIRD:2022,CMB-S4:science2016}, will be able to place stringent constraints on the speed of GWs in these Horndeski models. Finally, we note that \cite{Raveri:2019mxg} did not find a qualitative difference between reconstructed Horndeski models with zero and non-zero $\gamma^{\mathrm{MG}}_3(a)$ when confronting models with current cosmological data\footnote{Their data sets include CMB (Planck), weak lensing (CFHTLenS), BAO (6dFGS and SDSS) and type Ia supernovae (Pantheon).}.

\subsection{Stability conditions in EFTDE and \code{EFTCAMB}}\label{sec:EFTCAMB_stability}

\code{EFTCAMB} further allows user to impose a set of consistency checks on the EFT functions in order to ensure that the EFTDE models being considered and evaluated meet the theoretical stability conditions \cite{Hu:2013twa}.
These so-called \emph{viability conditions} \cite{Hu:2014oga}, or rather \emph{viability priors} in the context of cosmological inference \cite{Raveri:2014cka}, include
\begin{enumerate}
    \item Physical stability: the EFTDE theory must have a background stable to perturbations. In other words, the background must be free from ghost and gradient instabilities. The former corresponds to the situation where the model has a negative kinetic energy; the latter refers to the scenario where the squared sound speed is negative in some background regions. \cite{Bellini:2014fua} (see Eqs.~(42)-(51) of \cite{Hu:2014oga} for details).
    \item Mathematical stability: the EFTDE theory must have a well-defined $\pi$-field equation with no fast exponential growing modes of perturbations, as well as well-defined equations for tensor perturbations (see Eq.~(52) of \cite{Hu:2014oga} for details).
    \item Additional, model-specific stability: For Horndeski models, this enforces $w(a)\leq-1/3$ at all time. Specifically to this work, this condition is automatically guaranteed as we consider only the case of a constant $w=-1$.
\end{enumerate}
Generally speaking, the set of physical stability conditions is more restrictive than the mathematical ones. Further, the mathematical stability conditions implicitly assume that a) the $\pi$-field equation decouple from other field equations and b) its time-dependent coefficients evolve slowly (with time); these conditions are approximate and model-dependent. Therefore, in this work we only impose the physical conditions. We have explicitly verified through a number of pilot runs that running \code{EFTCAMB} with only physical conditions versus both physical and mathematical conditions does not qualitatively affect the range of Horndeski models successfully evaluated by \code{EFTCAMB}, hence the principal results of our work. 

\subsection{Growth prediction in Horndeski models}\label{sec:growth_in_Horndeski}

In order to draw connections between Horndeski models and observational data, we will focus on the prediction of each Horndeski model for the parameter combination $f(z)\sigma_8(z)\equiv\fs(z)$.
This quantity plays a central role in describing galaxy peculiar velocities and redshift-space distortions; it is thus an excellent meeting place between observations and theories of modified gravity. For each theoretical model under consideration, we compute the exact $\fs(z)$ using \code{EFTCAMB} in bins of redshift. In this paper, we follow the convention in \cite{Planck:2015fie,Planck:2018vyg} and define $\fs(z)$ through
\begin{equation}
    \fs (z) \equiv \frac{\left[ \sigma_8^{(vd)} \left(z \right) \right]^2}{\sigma_8^{(dd)} \left( z \right)},
    \label{eq:CAMB_fsigma8}
\end{equation}
where $\sigma_8^{(vd)}$ is the amplitude of (total) matter fluctuations obtained from the matter velocity-density (cross-)correlation function, while $\sigma_8^{(dd)}\equiv\sigma_8$ is that same quantity obtained from the matter density-density (auto-)correlation function.
Specifically,
\begin{equation}
\left[\sigma^{(xx)}_8(z)\right]^2 = \int \frac{dk}{k} W^2_{\mathrm{TH}}(k,R=8\hinvmpc) T_x(k,z) T_x(k,z) P_{\mathcal{R}}(k),
    \label{eq:CAMB_sigma8}
\end{equation}
where $x$ denotes either the $v$ or $d$ component, $W_{\mathrm{TH}}(k,R=8\hinvmpc)$ is the Fourier transform of the spherical top-hat window function of radius $R=8\hinvmpc$, $T_x(k)$ is the transfer function of the $x$ component, and $P_{\mathcal{R}}(k)$ is the power spectrum of primordial adiabatic perturbations.

Dividing this $\fs(z)$ by the value of $\sigma_8(z)$ (also calculated by \code{EFTCAMB}), gives the theoretically predicted growth rate $f(z)$ of each Horndeski model, which will then be fit with the formula in \refeq{gamma_parameterization} with a specified functional form of $\gamma(z)$. For all Horndeski models considered in this work, \refeqs{CAMB_fsigma8}{CAMB_sigma8} or \refeq{growth_rate} yields the same numerical result and quantitative conclusion within the scales of interest, $k\simeq 0.01-0.1 \, \hmpcinv$.
This conclusion naturally follows under the assumption that growth rate is scale-independent. Even though this assumption may not hold in more generic Horndeski and modified gravity models (see e.g.~\cite{Silvestri:2013,Baker:2014}), it holds up rather well for Horndeski models we consider here, in particular within the scales probed by Stage IV and V surveys, i.e. $k\simeq 0.01-0.1 \, \hmpcinv$. Within that range of $k$ and each of the Horndeski models considered in this work, $\fs(z)$ only varies within sub-percent level at any given $z$. We further illustrate and discuss this point in \refapp{scaledep}.

\section{Testing growth parameterizations in Horndeski models}\label{sec:fitting_form}

Our aim is to statistically chart a broad range of functional forms of $\gamma(z)$, but only for Horndeski models that are compatible with current observational constraints. To do so, we first identify the sub-space of Horndeski theories in which models are \emph{both} stable and compatible with current constraints on $\fs(z)$. Detailed description about how we carry this out can be found in \refapp{scan}. 

\begin{figure}
    \centering
    \includegraphics[width=0.99\linewidth, center]{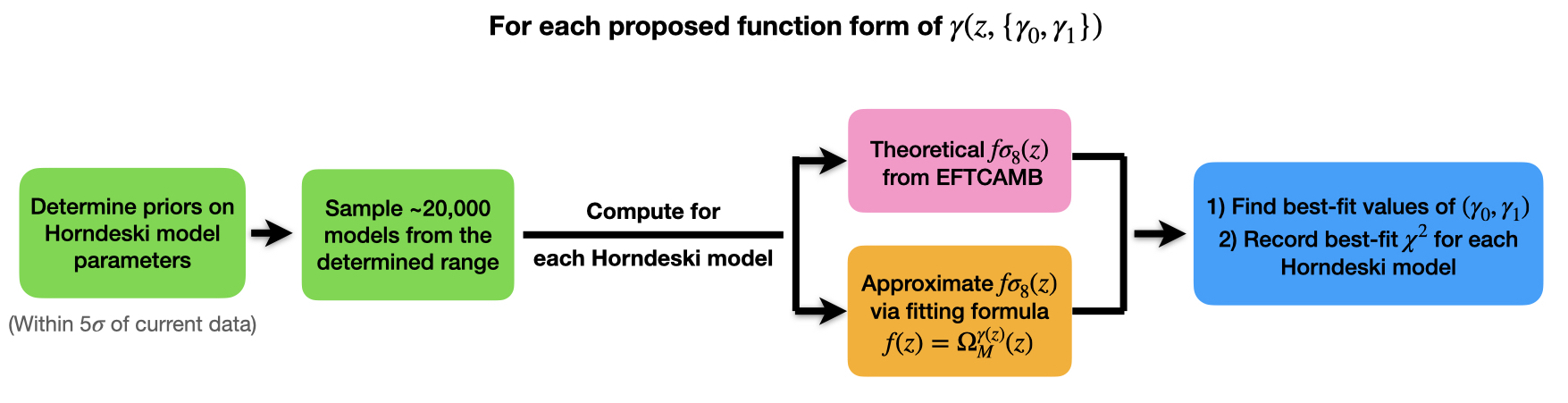}
    \caption{Flowchart describing how we sample Horndeski models from EFTDE theory space, evaluate the models and test the fitting functions for $\gamma(z)$ against their predictions for $\fs(z)$.}
    \label{fig:flowchart}
\end{figure}

After we have determined a sub-space of Horndeski theories compatible with current data, we then follow the procedure outlined here:
\begin{enumerate}
    \item We sample this theory sub-space, i.e.\ randomly draw Horndeski models from the sub-space and calculate the theoretical prediction for $\fs(z)$ by each model; this is described in \refsec{sample}.
    \item For each model, we quantify the goodness-of-fit between the $\fs(z)$ computed using various proposed fitting formulae for $\gamma(z)$ and the actual theory prediction. To compute the fit, we use the forecast constraints on $\fs(z)$ from Stage IV and V surveys; this is discussed in \refsec{fit}.
\end{enumerate}
Finally, we compare the goodness-of-fits and identify the best functional form for $\gamma(z)$. \reffig{flowchart} depicts the entire procedure.

\subsection{Sampling and evaluating Horndeski models}\label{sec:sample}

Following the preliminary runs and cuts described in \refapp{scan}, we identify a particular sub-space of Horndeski theories where models that are both stable and consistent with current data constraints on $\fs$:

\begin{equation}
\begin{aligned}
    \OmgZero &\in \mathcal{U}[0.0, 0.1], \; &&s_0 \in \mathcal{U}[0,3]\\[0.2cm]
    \gone &\in \mathcal{U}[0.0,0.7], \; &&s_1 \in \mathcal{U}[-3,3],\\[0.2cm]
    \gtwo &\in \mathcal{U}[-1.0,0.0], \; &&s_2 \in \mathcal{U}[0,3],\\[0.2cm]
    \gthree &\in \mathcal{U}[0.0, 1.0], \; &&s_3 \in \mathcal{U}[1,3].
      \label{eq:eft_params_range_final}
\end{aligned}
\end{equation}
\refeq{eft_params_range_final} specifies our priors on EFTDE parameters from which we sample Horndeski models in our main analysis. These priors are in broad agreement with the corresponding posteriors reported in \cite{Frusciante:2018jzw}.

From the Horndeski priors in \refeq{eft_params_range_final} and cosmological priors in \reftab{fiducial_values}, we first draw and evaluate a sample of $\sim$20,000 Horndeski models using \code{EFTCAMB}. Of these, 19,908 models pass the stability conditions imposed within \code{EFTCAMB} (see \refsec{theory}).
For each model successfully evaluated by \code{EFTCAMB}, we use its prediction of $\fs(z)$ to recompute the quantity $\chicurrent$ given in \refeq{chisq_current} and reject those with $\chicurrent>5\sigma$ (same cut as in \refapp{scan}).
We thereby end up with a final set of 18,543 viable Horndeski models  and we only use \emph{these} models to test our fitting formulae and identify the most accurate one.

\def\arraystretch{1.2}
\setlength\tabcolsep{0.1cm}
\begin{table}[t]
    \caption{Fiducial values of cosmological parameters and priors on them used in our sampling of Horndeski models. They closely follow the $\Lambda$CDM best-fit values and 68\% in the Planck 2018 analysis \cite{Planck:2018vyg}.}
  \begin{center}
    \begin{tabular}{c|c|c}
      
      \hline\hline 
      \textbf{Parameter} & \textbf{Fiducial value} & \textbf{Prior distribution} \\

      \hline\hline
      $\Omega_b h^2$ & 0.022383 & $\mathcal{N}(0.022383,0.00015)$ \\
      $\Omega_c h^2$ & 0.12011 & $\mathcal{N}(0.12011,0.0012)$  \\
      $H_0$ & 67.32 & $\mathcal{N}(67.32,0.54)$   \\
      \hline
      $w$ & $-1.0$ & \multirow{4}{*}{Fixed} \\
      $A_s$ & 2.086 $\times 10^{-9}$ & \\
      $n_s$ & 0.9666 &   \\
      $\tau$ &0.0543 &   \\
      \hline\hline 
    \end{tabular}
  \end{center}
  \label{table:fiducial_values}
\end{table}

\def\arraystretch{1.1}
\setlength\tabcolsep{0.15cm}
\begin{table}[t]
    \caption{Current measurements of $f \sigma_8$ and errors at different redshifts. The data include the 6dF Galaxy Survey (6dFGS), peculiar velocities of type Ia supernovae (SNIa), Galaxy and Mass Assembly (GAMA), the WiggleZ Dark Energy Survey, the Baryon Oscillation Spectroscopic Survey (BOSS), the extended Baryon Oscillation Spectroscopic Survey (eBOSS) and the VIMOS Public Extragalactic Redshift Survey (VIPERS).}
  \begin{center}
    \begin{tabular}{c|c|c|c}
      
      \hline\hline 
      \textbf{Redshift} & \boldsymbol{$f \sigma_8$} & \boldsymbol{$\sigma_{f \sigma_8}$} & \textbf{Survey/Probe} \\

      \hline\hline
      0 & 0.418 & 0.065 & 6dFGS \cite{Johnson:2014} \\ 
      \hline
      0 & 0.40 & 0.07 & SNIa \cite{Turnbull:2011ty} \\
      \hline
      0.067 & 0.423 & 0.055 & 6dFGS \cite{Beutler:2012px} \\
      \hline
      0.18 & 0.44 & 0.06 & \multirow{2}{*}{GAMA \cite{Blake:2013nif}} \\
      0.38 & 0.44 & 0.06 & \\
      \hline
      0.22 & 0.42 & 0.07 & \multirow{4}{*}{WiggleZ \cite{Blake:2011rj}} \\
      0.41 & 0.45 & 0.04 & \\
      0.60 & 0.43 & 0.04 & \\
      0.78 & 0.38 & 0.04 & \\
      \hline
      0.38 & 0.482 & 0.053 & \multirow{3}{*}{BOSS \cite{BOSS:2016psr}} \\
      0.51 & 0.455 & 0.050 & \\
      0.61 & 0.410 & 0.042 & \\
      \hline
      0.57 & 0.441 & 0.044 & BOSS RSD \cite{Samushia:2013yga} \\ 
      \hline
      0.15 & 0.53 & 0.16 & \multirow{6}{*}{eBOSS \cite{Alam:2016}} \\
      0.38 & 0.500 & 0.047 & \\
      0.51 & 0.455 & 0.039 & \\
      0.70 & 0.448 & 0.043 & \\
      0.85 & 0.315 & 0.095 & \\
      1.48 & 0.462 & 0.045 & \\ 
      \hline
      0.80 & 0.47 & 0.08 & VIPERS \cite{delaTorre:2013rpa}\\
      \hline\hline 
    \end{tabular}
  \end{center}
  \label{table:current_data}
\end{table}

\subsection{Testing the functional forms for \texorpdfstring{$\gamma(z)$}{Lg}}\label{sec:fit}

In this section, we find the most accurate two-parameter description of the growth index that fits Horndeski models, assuming future $\fs$ data. 

\begin{figure}
    \centering
    \includegraphics[width=0.85\linewidth, center]{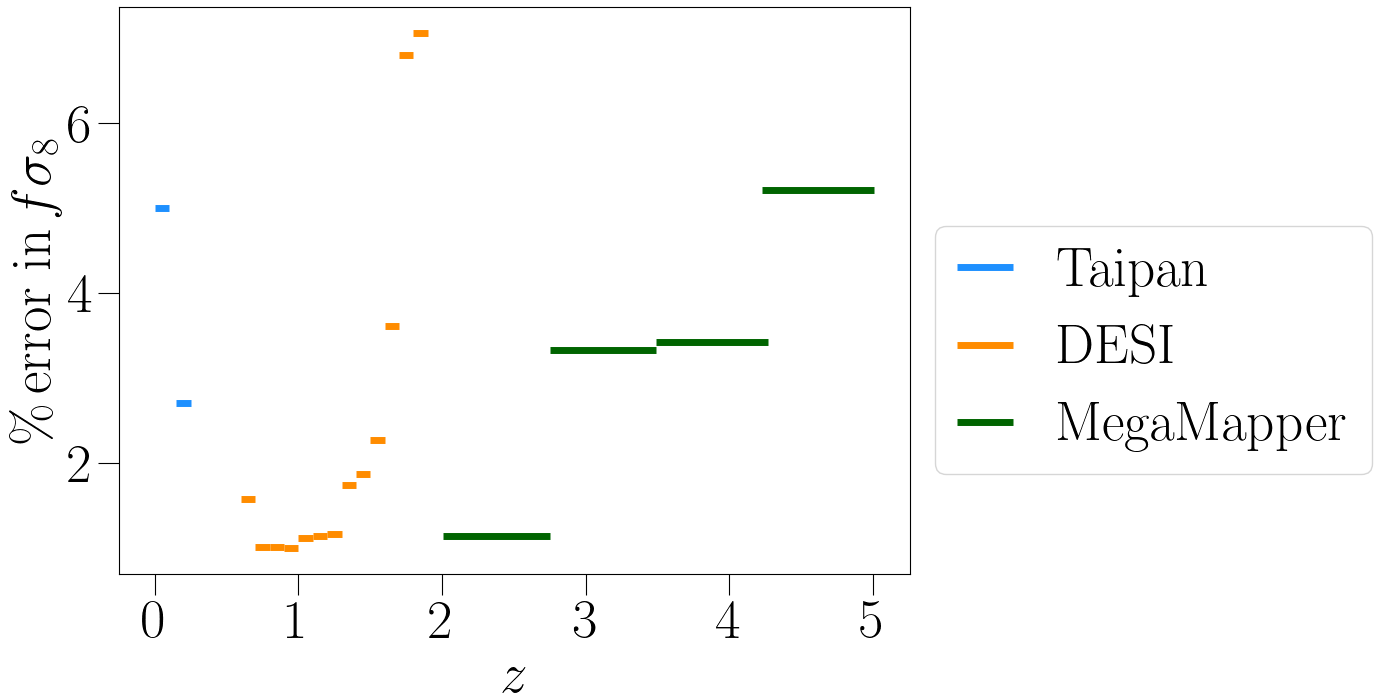}
    \caption{Future $\fs(z)$ error forecasts that we use to assess the accuracy of the fitting formulae for the growth rate. The width of each line segment indicates the size of the redshift bin while the height shows the magnitude of the forecast error.}
    \label{fig:error_bar}
\end{figure}

The final fitting formula must be sufficiently accurate even for future measurements of structure growth in the coming years and decades. We therefore assume optimistic constraints given by future data considered in this work. In this way, we impose a high burden of proof for any proposed fitting formula.
Specifically, we choose the measurements of, and constraints on, $f\sigma_8(z)$ from several future surveys that together cover a redshift span up to $z_{\rm max} = 5$, as shown in Figure~\ref{fig:error_bar} and listed in \reftab{future_data}. In the low-redshift region, we adopt forecasted error bars based on the Taipan Galaxy Survey \cite{Taipan:whitepaper2017}, assuming a $5\%$ error at $z = 0.05$ and a $2.7\%$ error at $z = 0.2$. In the intermediate redshift range of $0.65 < z < 1.85$, we adopt the DESI forecasts \cite{DESI-PartI:2016} where errors are in bins of size $\Delta z = 0.1$ as given in \reftab{future_data}. In the high redshift region, we use forecasts from MegaMapper\footnote{The $\fs$ constraint forecast for MegaMapper was obtained through a joint fit to $\{\Omega_c,\Omega_b,h,\log A_s\}$, marginalizing over galaxy bias and nuisance parameters.} where errors are in four bins of size $\Delta z = 0.75$.  

\def\arraystretch{1.1}
\setlength\tabcolsep{0.15cm}
\begin{table}[t]
    \caption{Constraints on $f \sigma_8$ from future surveys that cover a redshift range of up to $z_{\rm max} = 5$. This is also visualized in \reffig{error_bar}.}
  \begin{center}
    \begin{tabular}{c|c|c}
      
      \hline\hline 
      \textbf{Redshift} & \% \textbf{Error in} \boldsymbol{$\fs(z)$} & \textbf{Survey/Probe} \\
      
      \hline\hline
      0.05 & 5 & \multirow{2}{*}{Taipan \cite{Taipan:whitepaper2017}} \\
      0.2 & 2.7 &  \\
      \hline
      0.65 & 1.57 & \multirow{13}{*}{DESI \cite{DESI-PartI:2016}} \\
      0.75 & 1.01 & \\
      0.85 & 1.0 & \\
      0.95 & 0.99 & \\
      1.05 & 1.11 & \\
      1.15 & 1.14 & \\
      1.25 & 1.16 & \\
      1.35 & 1.73 & \\
      1.45 & 1.87 & \\
      1.55 & 2.27 & \\
      1.65 & 3.61 & \\
      1.75 & 6.81 & \\
      1.85 & 7.07 & \\
      \hline
      2.38 & 1.13 & \multirow{4}{*}{MegaMapper} \\
      3.12 & 3.33 & \\
      3.88 & 3.42 & \\
      4.62 & 5.21 & \\
 
      \hline\hline 
    \end{tabular}
  \end{center}
  \label{table:future_data}
\end{table}

For every Horndeski model (generated following the protocols described in \refapp{scan}), we assume future data centered on the predictions of that model, with errors representative of Stage IV and V surveys shown in \reftab{future_data}. We then fit this simulated data with a number of distinct two-parameter fitting formulae for $\gamma(z)$.  To perform each fit, we employ the \texttt{iminuit} optimization package to find the best-fit values of the two fitting-formula parameters, $\gamma_0$ and $\gamma_1$ that minimizes $\chifit$, which is defined as 
\begin{equation}
\chifit\equiv \sum_i\frac{[(\fs)^{\rm fit}(z_i)-(\fs)^{\rm model}(z_i)]^2}{(\sigma^{\rm future\, data}_i)^2}. 
\label{eq:chifit}
\end{equation}
Here $(\fs)^{\rm model}(z)$ is the value obtained directly from \code{EFTCAMB} following the definition in \refeq{CAMB_fsigma8}; in $(\fs)^{\rm fit}(z)$, $f(z)$ was calculated by each of the two-parameter parameterizations listed in \reftab{different_fitting_formula} and $\sigma_8(z)$ obtained from \code{EFTCAMB} following the definition in \refeq{CAMB_sigma8}; $\sigma^{\rm future\, data}_i$ is the error on the $i$-th future measurement at redshift $z_i$, both of which are given in \reftab{future_data}. Finally, we select the fitting formula of $\gamma(z)$ that gives the best goodness-of-fit across all sampled Horndeski models. 

\section{Results}\label{sec:results}

We now present the two principal results of this paper.
In \refsec{winner}, we show the performance of different parameterizations of $\gamma(z)$ in their fit to theoretical predictions of $\fs$ in Horndeski models and identify the best fitting formula. In \refsec{data_constraints} we constrain the parameters of the best fitting function using current cosmological data.

\subsection{And the winner is...}\label{sec:winner}

\begin{figure}
    \centering
    \includegraphics[width=0.8\linewidth]{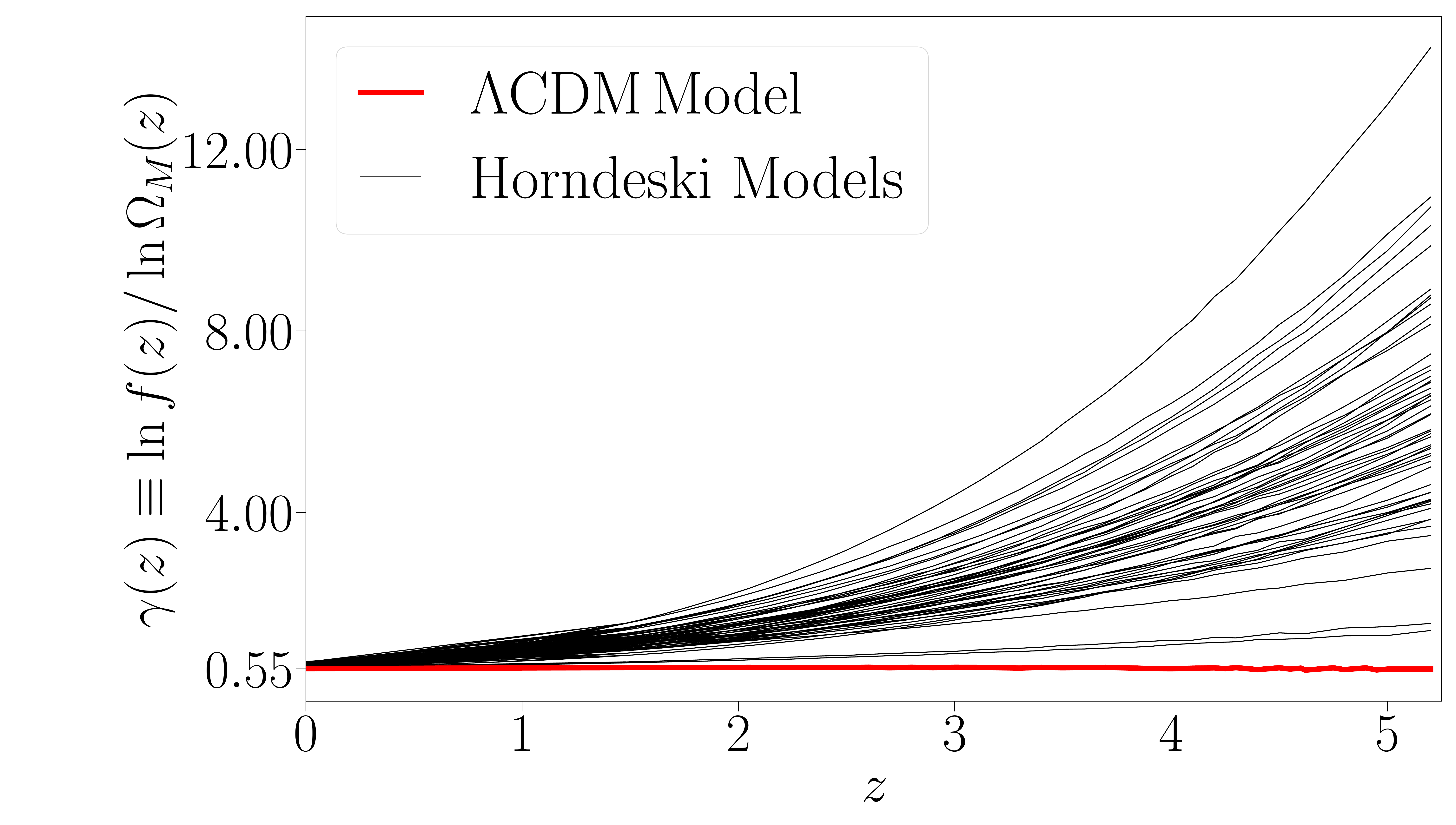}
    \caption{The exact redshift-dependent growth index, $\gamma(z) \equiv \ln{f(z)} / \ln{\Omega_m (z)}$, numerically evaluated for 50 randomly selected Horndeski models from our prior out to $z_{\rm max} \simeq 5$. The red nearly horizontal line shows the exact $\gamma(z)$ for the $\Lambda$CDM model. These results demonstrate that one needs a nonlinear multi-parameter parameterization to capture the features of growth index at high redshift in modified gravity. They also motivate functional forms for our trial fitting functions in \reftab{different_fitting_formula}.}
    \label{fig:lnf}
\end{figure}

To first get a sense of what kind of $\gamma(z)$ is typically predicted by Horndeski models, we numerically compute the redshift-dependent growth index for a limited number of models. Specifically, we evaluate the true growth index
\begin{equation}
    \gamma(z)\equiv \frac{\ln{f}(z)}{\ln\Omega_M (z)}
    \label{eq:gamma_z}
\end{equation}
given $f(z)$ and $\Omega_M(z)$ in that model. \reffig{lnf} shows the exact $\gamma(z)$ computed from \refeq{gamma_z} in 50 Horndeski models randomly selected from our prior. To guide the eye we also plot the $\Lambda$CDM growth index which, as expected, is very well approximated by $\gamma(z)\simeq 0.55$. The general behavior of $\gamma(z)$ in Horndeski models at $z\gtrsim 1$ can be easily understood from \refeq{gamma_z}: as $z$ increases, $\Omega_M(z)\rightarrow 1$ in cosmological models without early dark energy (which is true for all models considered in this paper). Therefore, for any given (Horndeski) model, departures of the growth rate $f(z\gtrsim 1)$ from the $\Lambda$CDM prediction are generally associated with relatively large fluctuations in the growth index simply because the latter is the exponent of $\Omega_M(z)$, which is close to unity. Overall, the results in \reffig{lnf}  not only show a clear redshift evolution of the growth index expected in these models but also demonstrate that this redshift dependence is fairly featureless. 

The results in \reffig{lnf} motivate our selection of specific functional forms of the growth index and \reftab{different_fitting_formula} enumerates these functions. In addition to the constant growth index and (for pure simplicity) the one going linearly with $z$, we also try several other forms that contain simple polynomials in redshift, as well as simple logarithmic or exponential terms. Since we would like to parameterize $\fs(z)$ up to $z \simeq 5$ where Stage IV and V data will constrain growth, trends in \reffig{lnf} emphasize that we need a nonlinear redshift-dependent parameterization, such as those suggested in \reftab{different_fitting_formula}.


The success (or failure) of each parameterization in fitting theoretical predictions of Horndeski models is further reported in \reftab{different_fitting_formula}. For each fitting function, we summarize the statistics of the quantity $\chifit$, defined in \refeq{chifit}, measured for our set of $\sim$18,000 Horndeski models. The summary is provided by two statistical measures: the median of the $\chifit$ values, and the 95-th percentile (i.e.\ the upper bound of 95\% of values of $\chifit$). Since the theoretical data vector calculated by \code{EFTCAMB} is noiseless, a perfect fit of a fitting formula to true $\fs(z)$  of a theory model will have $\chifit=0$; this explains the generally small chi-squared values in \reftab{different_fitting_formula}. In general, we find that the distribution of the $\chifit$ values  has a heavy tail in each instance; this explains why the 95\% upper bounds are typically much larger than the corresponding medians in \reftab{different_fitting_formula}.
The presence of the heavy tails reflects the improvement of $\fs$ constraints going from Stage III to Stage IV and V surveys: our model selection cut, \refeq{chisq_current}, is only concerned with current constraints on $\fs$, while our comparison by \refeq{chifit} is concerned with forecast constraints for Stage IV and V surveys.

\begin{table}[ht]
    \caption{Proposed fitting functions and the statistics of their fit to our sample of $\sim$18,000  Horndeski models.}
    \begin{center}
        \begin{tabular}{c|c c}
          \hline
            \textbf{Fitting function} & \multicolumn{2}{c}{\textbf{Best-fit} \boldsymbol{$\chi^2$}} \\ 
            \textbf{for} \boldsymbol{$\gamma(z)$} & \textbf{Median} & \textbf{95\% Percentile} \\
         \hline
         $\gamma_0$ & 1.16 & 36.6 \\
          \hline
           $\gamma_0 + \gamma_1 z$ & 0.046 & 4.00  \\
          
        $\gamma_0 + \gamma_1 z^2$ & 0.11 & 2.48 \\
        
        \hline

        $\gamma_0 + \gamma_1 z / (1+z)$ & 0.22 & 14.7 \\
        
        \cellcolor{red!25}$\gamma_0 + \gamma_1 z^2 / (1+z)$ & 0.028 & 1.08 \\
        
        $\gamma_0 + \gamma_1 z^3 / (1+z)$ & 0.26 & 6.58 \\
        
        \hline

        $\gamma_0 + \gamma_1  \exp(z) $ & 0.28 & 6.91 \\
        
        $\gamma_0 + \gamma_1 z^3 \exp(-z) $ & 0.10 & 2.63 \\
        
          \hline
         
        \end{tabular}
    \end{center}
    
    \label{table:different_fitting_formula}
\end{table}

As shown by the highlighted cell in \reftab{different_fitting_formula}, the best fitting formula for the growth index in the redshift range of $0 < z < 5$ is 
\begin{equation}
    \gamma(z) = \gamma_0 + \gamma_1 \frac{z^2}{1+z}.
    \label{eq:best_formula}
\end{equation}
This two-parameter fitting formula fits the future data with a median $\chi^2_{\rm fit}$ of 0.028, which is about 40 times smaller than the median $\chifit$ with the constant growth index $\gamma(z) = \gamma_0$.  Furthermore,  we find that with the new fitting formula from \refeq{best_formula}, the maximum deviation in $\fs(z)$ at any redshift between the fitting formula's approximation and Horndeski theory's true value for $\fs(z)$ has a median of 0.4\% when averaged over all models. When using the traditional one-parameter growth index, $\gamma(z) = \gamma_0$, the median of maximum differences per model is 2.5\%.
Our two-parameter, redshift-dependent fitting formula therefore approximates the theoretical predictions of $\fs(z)$ in Horndeski theories about six times better than the one-parameter, constant-$\gamma$ case, leading to an improvement of $\sim 40$ times in $\chi^2$.

Several other fitting functions in \reftab{different_fitting_formula}, also do a good job, in particular $\gamma(z) = \gamma_0 + \gamma_1 z$ comes close in the median, but falls short in fitting models near the tail; however, none are as good as the form in \refeq{best_formula}.

The result that the median $\chifit \ll 1$ with the best fitting function is very encouraging, as it implies that the contribution of the inaccurate fitting function to the bias in cosmological parameters will be subdominant. Specifically, our finding that $\chifit \ll 1$ implies that even the best-determined direction in parameter space will be biased by $\ll$ 1$\sigma$ in our optimistic-data case.

\begin{figure*}[t]
    \centering
    \includegraphics[width=\linewidth]{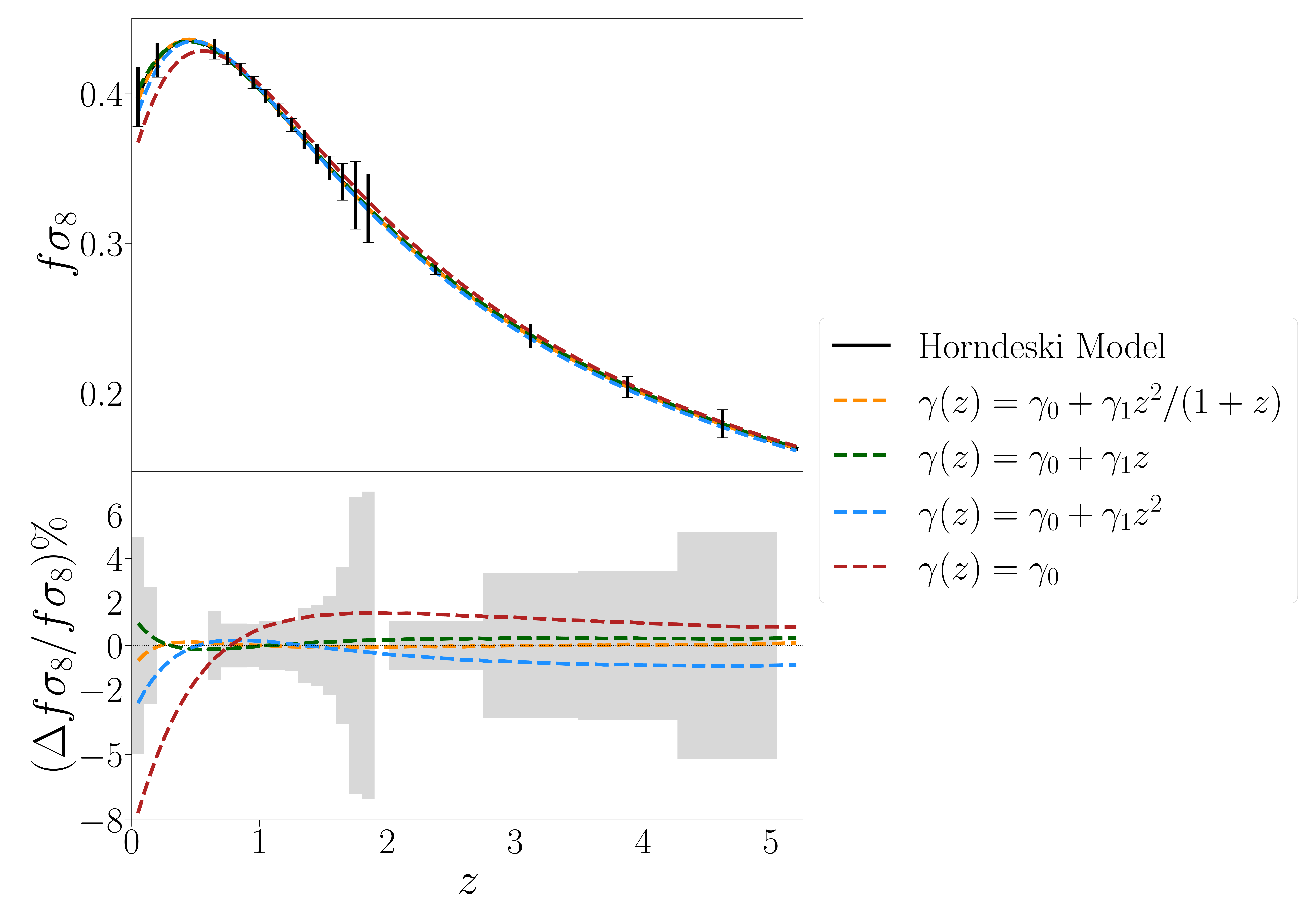}
    \caption{\textit{Top panel}: The $\fs(z)$ of an example Horndeski model (black curve, based on parameters $\OmgZero = 0.074$, $s_0 = 2.33$, $\gone = 0.15$, $s_1 = -0.57$, $\gtwo = -0.92$, $s_2 = 1.48$, $\gthree = 0.75$ and $s_3 = 1.43$) computed by \code{EFTCAMB} with error bars forecast for future surveys from \reftab{future_data}: Taipan, DESI and MegaMapper. We also show the best-fitting $\fs(z)$ for three fitting formulae for the growth index $\gamma(z)$, as well as the $\fs$ fit from a constant growth index, $\gamma(z) = \gamma_0$. \textit{Bottom panel}: Relative difference between the true $\fs(z)$ and the best-fit results of each fitting formula shown in the top panel. The shaded grey area shows the forecast statistical errors associated with each survey.}
    \label{fig:diff_functions_to_Horndeski}
\end{figure*}

In \reffig{diff_functions_to_Horndeski}, we further showcase the performance of a few proposed fitting formulae on one randomly selected Horndeski model. It is evident that the prize-winning fitting form in \refeq{best_formula} is the best of the fitting functions shown. We also observe that the best fitting function does a good job both at $z\ll 1$ and at $z>1$; both of these ranges are required to be accurately fit for the $(\gamma_0, \gamma_1)$ description to be a useful tool for the Stage IV and V surveys.

\subsection{Constraint on \texorpdfstring{$\gamma(z)=\gamma_0+\gamma_1z^2/(1+z)${ f}} from current data}\label{sec:data_constraints}

In the previous section, we have proposed and validated \refeq{best_formula} as a new fitting function of the growth index $\gamma(z)$ for future surveys. Using current cosmological data, we now demonstrate the applicability of this formula in consistency tests of general relativity and flat \LCDM.

\begin{figure*}
    \centering
    \includegraphics[width=0.8\linewidth]{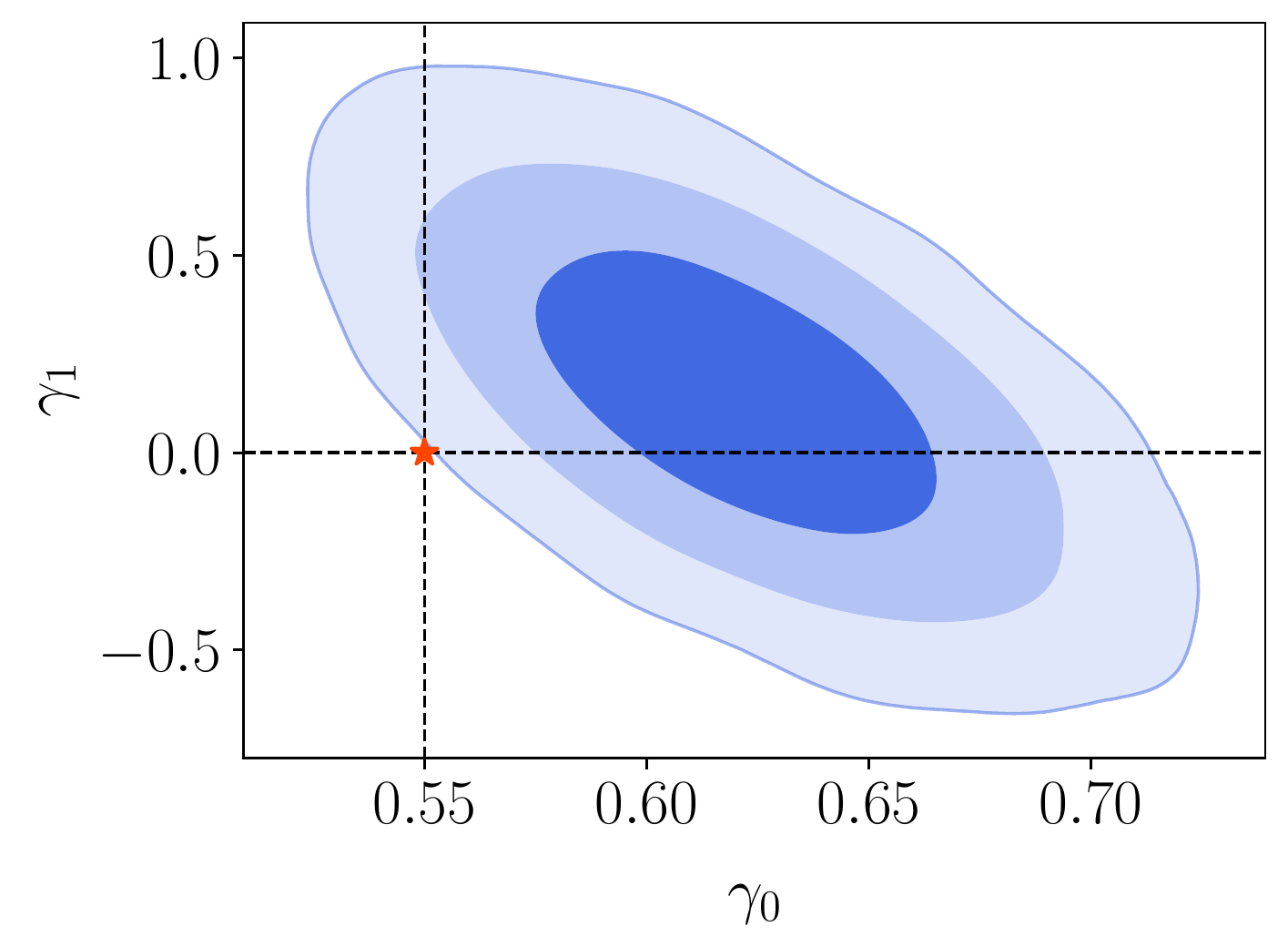}
    \caption{Constraints from current cosmological data on parameters $(\gamma_0,\gamma_1)$ in the growth-index parameterization $\gamma(z)=\gamma_0+\gamma_1z^2/(1+z)$. Contours in the $\gamma_0-\gamma_1$ 2D plane represent (68\%, 95\%, 99.73\%) of the posterior volume.}
    \label{fig:gamma0_gamma1_constraints}
\end{figure*}

Building upon the work in \cite{Nguyen:2023}, here we constrain the growth index $\gamma(z)$, specifically $(\gamma_0,\gamma_1)$ in \refeq{best_formula}, from a combination of large-scale structure and CMB data sets: measurements of $\fs$ through peculiar velocities and redshift-space distortions\footnote{Fig. 2 of \cite{Nguyen:2023} shows these $\fs$ measurements and their error bars.} \cite{Beutler:2012px,Huterer:2016,Said:2020,Boruah:2020,Turner:2023,Blake:2011rj,Blake:2013nif,Howlett:2015,Okumura:2016,Pezzotta:2017,SDSS:DR16}, measurements of baryon acoustic oscillation (BAO) from the Six-degree Field Galaxy Survey (6dFGS; \cite{6DF:BAO_H0}) and the Sloan Digital Sky Survey (SDSS; \cite{SDSS:DR7,SDSS:DR12,SDSS:DR16}), 3x2pt correlation functions from the Year-1 analysis of Dark Energy Survey (DES-Y1; \cite{DESY1:3x2pt}), and CMB measurements from Planck 2018 \cite{Planck:2018vyg}. In this work, we additionally include the type Ia supernovae data sets and likelihoods from Pantheon \cite{Pantheon:2018} which however make very little difference in our final constraints on $(\gamma_0,\gamma_1)$. To obtain constraints on the growth-index and cosmological parameters (after numerically marginalizing over nuisance parameters), we use \code{cobaya}\footnote{\href{cobaya.readthedocs.io/en/latest/}{https://cobaya.readthedocs.io/en/latest/}}, which provides out-of-the-box access to most likelihoods for the aforementioned data sets, validated against their official analyses. The only exception is the $\fs$ likelihood for \cite{6dF:growth_sigma8,Huterer:2016,Said:2020,Boruah:2020,Turner:2023,Blake:2011rj,Blake:2013nif,Howlett:2015,Okumura:2016,Pezzotta:2017}, which we implement in \cite{Nguyen:2023}, assuming a Gaussian likelihood and a diagonal covariance.

Our implementation of the growth index largely follows that of \cite{Nguyen:2023}. Specifically, at any given redshift $z$, we re-scale the linear matter power spectrum as
\begin{equation}
  P(\gamma,k,z)=P(k,z=0)\,D^2(\gamma,z),
\label{eq:scaling_plin}
\end{equation}
where $D(\gamma,z)$ is numerically integrated from \refeqs{growth_rate}{gamma_parameterization}, and $P(k,z=0)$ is the fiducial linear matter power spectrum evaluated at $z=0$ which is specified by the standard set of cosmological parameters:
\begin{equation}
\{A_s,n_s,\Omega_c h^2,\Omega_b h^2,\tau,\theta_{\mathrm{MC}}\},
\label{eq:cosmological_params}
\end{equation}
where $A_s$ and $n_s$ are the amplitude and the spectral index of the primordial power spectrum, $\tau$ is the reionization optical depth, and $\theta_{\mathrm{MC}}$ is (an approximation to) the angular size of the sound horizon at recombination. We emphasize that this set of cosmological parameters is jointly constrained with $(\gamma_0,\gamma_1)$. When sampling with \code{cobaya}, we compute $P(k,z)$ using the cosmological Boltzmann solver \code{CAMB} \cite{Lewis:camb,Howlett:camb}.
We validate our implementation by reproducing, up to a high precision, the constraints on the standard cosmological parameters in the baseline analyses of Planck 2018 \cite{Planck:2018vyg} and DES-Y1 \cite{DESY1:3x2pt}.

Motivated by the fact that \refeq{gamma_parameterization} has, so far, been validated only for sub-horizon perturbations, we exempt the primary CMB anisotropies from the rescaling in \refeq{scaling_plin}. In other words, the growth index never directly affects the \emph{unlensed} CMB power spectra, but rather only the CMB lensing potential. Consequently, only the CMB lensing amplitude is sensitive to any change in the growth index\footnote{Note that, strictly speaking, $\gamma$ should also affect the integrated Sachs-Wolfe (ISW) effect but here we do not consider a separate ISW likelihood. For more details on the latter, see \cite{Carron:2022eum}.} $\gamma$.

For the cosmological parameters, we adopt the same priors as specified in the Planck 2018 baseline analysis \cite{Planck:2018vyg}, which considered flat \LCDM\ at fixed neutrino mass $\sum m_\nu=0.06$ eV. Priors on all nuisance parameters also follow those in the official analyses of the corresponding data sets. We choose uniform priors on the two growth-index parameters:
$\gamma_0 \in \mathcal{U}(0.0,2.0)$, $\gamma_1 \in \mathcal{U}(-1.0,1.0)$.

In \reffig{gamma0_gamma1_constraints} we present the  constraints in the $\gamma_0-\gamma_1$ plane, marginalized over all other cosmological and nuisance parameters.
Allowing for redshift evolution, which is effectively controlled by the parameter $\gamma_1$ in $\gamma(z)=\gamma_0+\gamma_1z^2/(1+z)$, we observe the expected degeneracy between $\gamma_1$ and $\gamma_0$ in that parameterization.
Specifically, we infer $\gamma_0=0.621\pm0.03$ and $\gamma_1=0.149\pm0.235$.
We find evidence for a disagreement with the standard cosmological model --- which predicts $(\gamma_0=0.55,\gamma_1=0)$ --- at approximately 99.8\% level (corresponding to about ``3.1-sigma'' in a two-tailed test of statistical significance). Our finding is therefore in good statistical agreement with the conclusion from the analysis in \cite{Nguyen:2023} which however assumed $\gamma=\mathrm{const.}$, i.e. no redshift evolution.

Our $\gamma_0$ constraint suggests that the growth rate of large-scale structure is suppressed recently --- with the onset of dark energy --- relative to the prediction by flat \LCDM\ and general relativity, while the $\gamma_1$ constraint implies no (strong) evidence of redshift evolution in the growth index.

\section{Summary and Conclusions}\label{sec:conclusion}

In order to squeeze out stringent constraints on the growth of structure from data, it will be crucial to have precise parameterizations of the evolution of the growth of structure, specifically with the goal to cleanly separate it from the background evolution. One such parameterization is $f(z)=\Omega_M(z)^\gamma$ with a constant growth index $\gamma$ which, while being highly accurate for dark-energy models close to $\Lambda$CDM, is no longer such for modified gravity. In this work, we have promoted $\gamma$ to a function of redshift, i.e. $\gamma(z)$. We have further identified and validated the best two-parameter fitting formula for $\gamma(z)$ that accurately describes the growth of structure across the landscape of Horndeski theories of modified gravity:
\begin{equation}
    f(z)=\Omega_M(z)^{\gamma_0+\gamma_1z^2/(1+z)}.
    \label{eq:final_result}
\end{equation}
We have explicitly shown that \refeq{final_result} fits the theoretical predictions of $\fs(z)$ by Horndeski models with typical errors at the sub-percent level, well within the precision that will be reached by Stage IV and Stage V surveys.

Further, as a demonstration, we have constrained the parameters of \refeq{final_result} using modern data from galaxy clustering, weak lensing, CMB, and type Ia supernovae. The result we obtained is in tension with the concordance cosmological model of $(\gamma_0, \gamma_1) = (0.55, 0)$ --- which was expected given such indications in our recent analysis which essentially assumed $\gamma(z)=\gamma_0$ \cite{Nguyen:2023}. Specifically, we have found evidence that $\gamma_0>0.55$, while the posterior of $\gamma_1$ peaks at positive values but is statistically consistent with zero.

We conclude that forthcoming data from ongoing and upcoming large-scale structure surveys \cite{Taipan:whitepaper2017,HETDEX:whitepaper2021,DESI:roadmap2022,PFS:whitepaper2014,Euclid:whitepaper2011,MegaMapper:whitepaper2022} will dramatically expand the redshift coverage and increase the precision of the growth-of-structure sector. This, in turn, will enable new opportunities to test the self-consistency of the standard cosmological model, and possibly detect deviations from general relativity from such measurements of the growth rate. Our new fitting formula should provide one reliable meeting point between data and theory.

\section{Data and software products}

The modified version of \code{CAMB} that implements our parameterization, $\gamma(z)=\gamma_0+\gamma_1z^2 / (1+z)$, is available at: \href{https://github.com/MinhMPA/CAMB_GammaPrime_Growth.git}{github.com/MinhMPA/CAMB\_GammaPrime\_Growth.git}. To obtain cosmological constraints from data using this parameterization, one can use the fork of \code{cobaya} at: \href{https://github.com/MinhMPA/cobaya.git}{github.com/MinhMPA/cobaya.git}.
To request the $f\sigma_8(z)$ data evaluated for the Horndeski models considered in this paper, kindly send a message to Y.W. and N.-M.N.

\section{Author Contributions}
Y.W and N.-M.N co-led this project and equally contributed.
Y.W.: co-led the project; methodology - conceptualization, pipeline development and validation; data product - validation and analyzation; result - validation and interpretation; writing - original draft, editing, final, and visualization.
N.-M.N.: co-led the project; methodology - conceptualization, pipeline development and validation; data product - production and analyzation; result - validation and interpretation; writing - original draft, editing, final and visualization.
D.H.:  methodology - conceptualization; result - validation and interpretation; writing - original draft, editing and final.

\section{Acknowledgements}

We thank Ilija Raki\'{c} for their initial collaboration on this project and Oliver Philcox for kindly providing the forecasts from MegaMapper on high redshift constraints to $\fs$. We are grateful to Eric Linder, Marco Raveri, and the referee for helpful discussions and valuable comments.
MN and DH acknowledge support from the Leinweber Center for Theoretical Physics, NASA grant under contract 19-ATP19-0058, DOE under contract DE-FG02-95ER40899. This research was supported in part through computational resources and services provided by Advanced Research Computing at the University of Michigan\footnote{\url{https://arc.umich.edu}} and the University of Michigan Research Computing Package\footnote{\url{https://arc.umich.edu/umrcp}}.

\appendix

\section{Determining sampling ranges for EFTDE and Horndeski parameters}\label{app:scan}

In this work, we need to sample and evaluate a large number of models from the Horndeski theory space. Therefore, it is crucial to identify the sub-space of Horndeski models that are stable and compatible with current observations, in particular, those of $\fs$ --- our main observable in the present study. We enforce this requirement following a two-step procedure:
\begin{enumerate}
\item First, we draw the standard cosmological parameters $\Omega_b h^2$, $\Omega_c h^2$, and $H_0$ in \refeq{pars} from the 1D marginal posteriors in Planck 2018 baseline analysis \cite{Planck:2018vyg}. We fix the rest of the background cosmological parameters, including the amplitude of the primordial power spectrum $A_s$ (at pivot wave number $k_{\rm piv}=0.05\ {\rm Mpc}^{-1}$), the scalar spectral index $n_s$, and the optical depth to reionization $\tau$, to the Planck 2018 baseline best-fit values (see first column of Tab. 1 in \cite{Planck:2018vyg}). For clarity, we summarize the parameter prior ranges and values in \reftab{fiducial_values}. Next, we draw the EFT parameters in \refeq{pars} from the following ranges

\begin{equation}
    \begin{aligned}
    \OmgZero &\in \mathcal{U}[0, 0.1], \; &&s_0 \in \mathcal{U}[0,3];\\[0.2cm]
    \gone &\in \mathcal{U}[0.0,1.0], \; &&s_1 \in \mathcal{U}[-3,3];\\[0.2cm]
    \gtwo &\in \mathcal{U}[-1.0,1.0], \; &&s_2 \in \mathcal{U}[-3,3];\\[0.2cm]
    \gthree &\in \mathcal{U}[0.0, 1.0], \; &&s_3 \in \mathcal{U}[-3,3],
      \label{eq:eft_params_range_original}
    \end{aligned}
\end{equation}

where $\mathcal{U}[a,b]$ denotes a uniform distribution between $a$ and $b$.

\item In the second step, we then choose to exclude cosmological models --- specified by the above cosmological parameters and EFT parameters --- that are disfavored by current data at $\geq5\sigma$. The current $\fs$ data that we use are shown in \reftab{current_data}. We define the goodness of fit to the theoretical model value as 
\begin{equation}
\chicurrent\equiv \sum_i\frac{[(\fs)^{\rm current\,data}(z_i)-(\fs)^{\rm model}(z_i)]^2}{(\sigma^{\rm current\,data}_i)^2}  
\label{eq:chisq_current}
\end{equation}
where $(\fs)^{\rm model}(z)$ is the value obtained from \code{EFTCAMB},  and $(\fs)^{\rm current\,data}$, $\sigma^{\rm current\,data}_i$, and $z_i$ are respectively the measurement, error, and redshift of current data, all of which are given in \reftab{current_data}.
Similar to \cite{Frusciante:2016xoj}, we typically find that --- although model stability can strongly depend on $\gone$ and $s_1$ --- model prediction (in our case, for $\fs$) does not.
With 20 $\fs$ measurements and six EFT parameters to be fit, we have $N_{\rm dof} = 14$ degrees of freedom. Assuming a normal distribution of individual $\fs$ measurements, keeping the models that are within $5\sigma$ from current measurements then requires $\chi^2 \leq 60$. 
\end{enumerate}

During this process, we also prune regions of the Horndeski parameter space where models largely fail the stability conditions specified in \refsec{EFTCAMB_stability}. We thereby end up with the prior ranges specified in \refeq{eft_params_range_final}.

\section{Scale dependence of growth}\label{app:scaledep}

Here we take a closer look at the scale dependence of the growth rate $f(z, k)$ specifically in the context of Horndeski models. As mentioned near the end of Sec.~\ref{sec:theory}, this scale dependence is not guaranteed to be negligible for growth in models beyond smooth dark energy. By using \code{EFTCAMB} we can straightforwardly investigate the effect by directly computing the linear growth as the ratio of the matter transfer functions at two different redshifts \begin{equation}
    D(z, k) = \frac{T(z, k)}{T(z=0, k)}
\end{equation}
and then numerically evaluating $f(z, k)$ from \refeq{growth_rate}.

\begin{figure*}
    \centering
    \includegraphics[width=\linewidth]{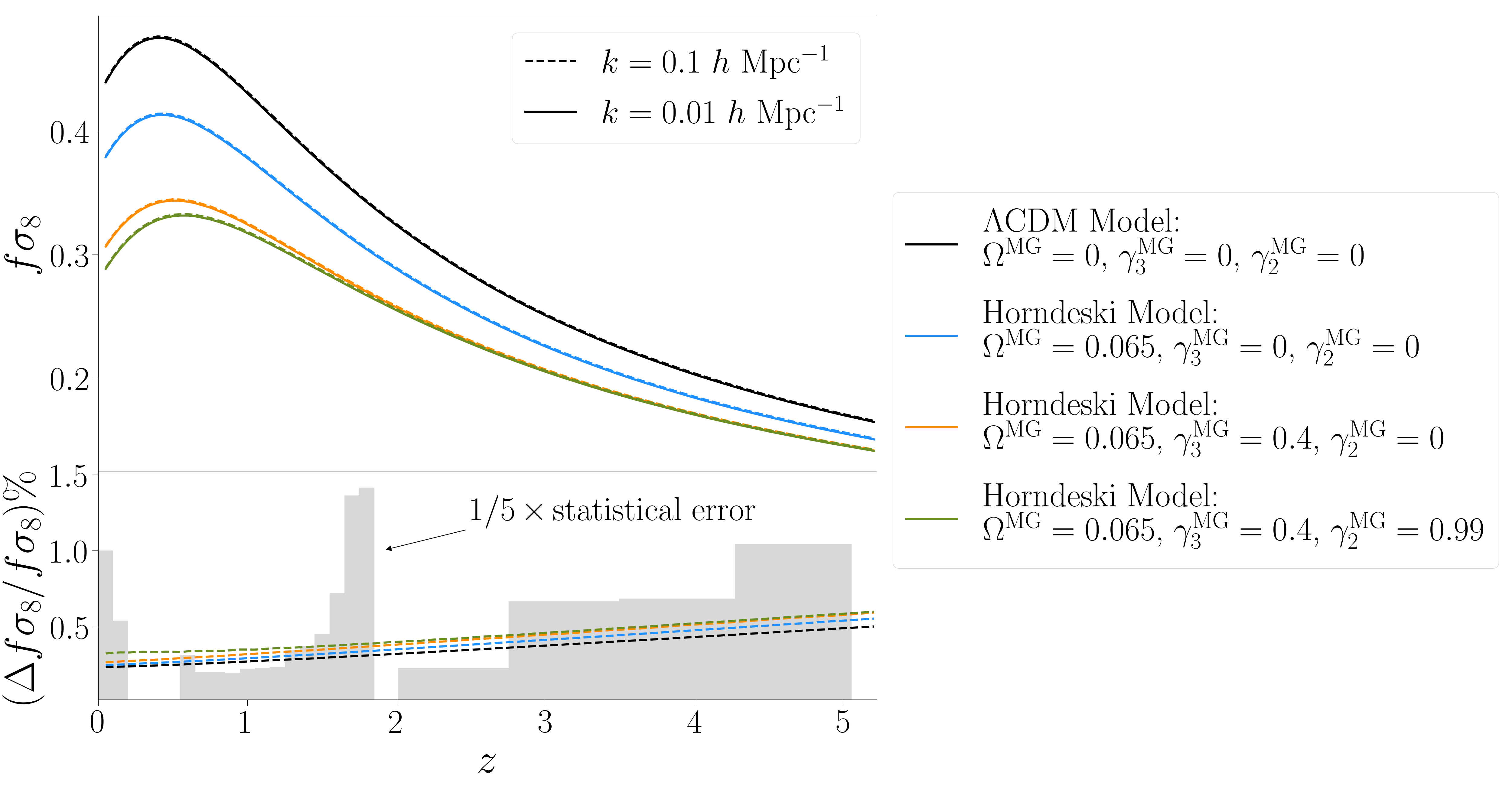}
    \caption{\textit{Top panel:} $\fs(z)$ of a $\Lambda$CDM and three Horndeski models calculated by \code{EFTCAMB} at scales $k = 0.1$ $h$ Mpc$^{-1}$ and $k = 0.01$ $h$ Mpc$^{-1}$. \textit{Bottom panel:} The percent difference between $\fs(z)$ of each Horndeski model calculated at these two scales. The shaded area illustrates the constraints on $\fs(z)$ by future surveys in terms of percent error in each redshift bin, which we use in determining the goodness-of-fit of each proposed fitting formula. For visualization purposes, we scale the errors down to $1/5$th of their true size in the bottom panel.}
    \label{fig:k_dependence}
\end{figure*}

We do find significant $k$ dependence of $f(z, k)$ near the horizon scale ($k\simeq 0.0001-0.001\hmpcinv$). However, note that most cosmological observations of large-scale structure come from smaller scales, roughly $k\simeq 0.01-0.1\hmpcinv$ (see, for example, Fig. 19 in the Planck legacy paper \cite{Planck:2018nkj}). Therefore, we do not need to take into account the scale dependence of $f(z)$ if we focus on this range of scales.
This is clearly demonstrated in \reffig{k_dependence} where we show the $k$ dependence of $\fs(z)$ on two scales,  $k= 0.01$ and $0.1\,\hmpcinv$, for a selection of a few Horndeski models from our priors (as well as for \LCDM).  When selecting the Horndeski models to showcase in Figure~\ref{fig:k_dependence}, we sequentially increase each EFT parameter to its largest value allowed by the priors specified in Eq.~(\ref{eq:eft_params_range_final}).

The maximum scale dependence that we observe in \reffig{k_dependence} is about 0.5\% --- much lower than the (statistical) errors in $\fs$ of all surveys considered in this work. We thus demonstrate that, for Horndeski models that deviate most from general relativity (and potentially have the most scale dependence) but are still within the our selected priors, the scale-dependent differences in $\fs(z)$ are well below $1\%$ and much smaller than stringent constraints from future surveys. 

We also quantitatively demonstrate the scale-independence of Horndeski models within our specified priors. Among $\sim$18,000 Horndeski models, we find that the difference between the values of $f(z, k)$ evaluated at $k = 0.01$ and $0.1\,\hmpcinv$ (across all models and all redshifts) has a median of 0.3\% and a 95\% percentile of 0.5\%, both at a sub-percent level.

Nevertheless, there are reasons why studying the scale dependence of the growth of structure is very interesting and should be pursued. First, modified-gravity models that are physically different from models in our Horndeski prior may lead to a much more significant scale dependence. Second, observations that probe larger spatial scales (say $k\sim0.001\,\hmpcinv$) might also be able to observe this scale-dependent behavior.

\bibliographystyle{JHEP} 
\bibliography{main}
\end{document}